\documentclass[12pt]{article}
\usepackage{bbm}
\usepackage{amssymb}
\usepackage{pifont}
\usepackage{mathrsfs}
\usepackage{amsfonts}
\usepackage{color}
\usepackage{footmisc}
\usepackage[numbers,sort&compress]{natbib}
\usepackage[colorlinks,linkcolor=blue,citecolor=blue,urlcolor=blue]{hyperref}

\newcommand{\omits}[1]{}

\begin{document}

\begin{center}
{\bf \LARGE On the difference between Poincar\'e\bigskip

and Lorentz gravity}
\bigskip\bigskip

{\large Jia-An Lu\footnote{Email: ljagdgz@163.com}}
\bigskip

School of Physics and Astronomy, Sun Yat-sen
University,\\ Guangzhou 510275, China
\bigskip

\begin{abstract}
The Poincar\'e invariance of GR is usually interpreted as Lorentz invariance plus
diffeomorphism invariance. In this paper, by introducing the local inertial
coordinates (LIC), it is shown that a theory with Lorentz and diffeomorphism
invariance is not necessarily Poincar\'e invariant. Actually, the energy-momentum
conservation is violated there. On the other hand, with the help of the LIC, the
Poincar\'e invariance is reinterpreted as an internal symmetry. In this formalism,
the conservation law is derived, which has not been sufficiently explored before.
\end{abstract}
\end{center}



\section{Introduction}
It is known from Utiyama's paper \cite{Utiyama} that GR can be viewed as a gauge
theory of the Lorentz group. Later the diffeomorphism invariance is identified to
the translational invariance, then GR is found to be a gauge theory of the Poincar\'e
group \cite{Kibble}. This interpretation of Poincar\'e invariance as a combination of Lorentz
invariance and diffeomorphism invariance becomes the standard one in the Poincar\'e
gauge theory of gravity \cite{Hehl76,Hehl95,Hehl13}. However, the diffeomorphism invariance
is not always equal to the translational invariance. For example, they are different
in the de Sitter (dS) gauge theory of gravity. In fact, the diffeomorphism symmetry
does not correspond to any conservation law in dS gravity \cite{Lu16}. Then one would
wonder whether the identification of diffeomorphism with translation in Poincar\'e
gravity is also problematic.

In this paper, we show that a theory with Lorentz invariance and diffeomorphism
invariance is not necessarily Poincar\'e invariant. We call such a theory the Lorentz
gauge theory of gravity. Just like the case of dS gravity, the diffeomorphism symmetry
does not correspond to any conservation law in Lorentz gravity. There exists only the
angular momentum (AM) conservation with respect to the Lorentz symmetry, in other words,
the energy-momentum (EM) conservation is absent.

Then how to interpret the Poincar\'e invariance when it is present?
The answer lies in the construction
of the Lorentz gravity. That is to introduce a vector field whose components may be called
the local inertial coordinates (LIC). The prototype of the LIC is Cartan's radius-vector
field \cite{Cartan,Hehl95}, which is $GL(n,\mathbb{R})$ covariant. The dS/AdS/Poincar\'e-covariant
LIC are first given by Guo \cite{Guo76}, West \cite{West} and Pilch \cite{Pilch}, respectively.
See also Refs. \cite{Lu13,Lu14} for a proof of their existence on an arbitrary spacetime.
With the help of the LIC, the Poincar\'e invariance can be interpreted as an
internal symmetry, just like the Lorentz symmetry. In this formalism, the fundamental variables
are the Poincar\'e connection and LIC. In the Lorentz gauge,
the LIC are fixed, and the Poincar\'e connection turns out to be a combination of the Lorentz
connection and tetrad field, which are the traditional variables for Poincar\'e gravity. For
completeness, we also compute the conservation law with respect to the Poincar\'e symmetry in
this formalism, which reduces to the ordinary one \cite{Kibble} in the Lorentz gauge. Although
Kawai \cite{Kawai} has attempted to do this, his result does not completely respect the spirit
of this formalism. The key point is that the Poincar\'e-covariant derivative has not entered
Kawai's conservation law, and thus the AM conservation has not been formulated into a neat form,
which will be given here.

The paper is organized as follows. In section 2, we construct the Lorentz gravity and derive the
conservation law. In section 3, the same thing is done except changing the Lorentz group to the
Poincar\'e group. In section 4, some remarks on the different formalisms and gauge groups
of gravity are presented.

\section{Lorentz gravity}
\subsection{Lorentz gravity from the gauge principle}
The Lorentz gravity is a gauge theory of the Lorentz group. Recall that in Weyl's
gauge theory, the gauge field is introduced to localize a global symmetry \cite{Weyl}.
First consider a classical field theory with global Lorentz invariance and diffeomorphism
invariance. The action integral and Lagrangian function of this matter field are as follows:
\begin{equation}\label{SM}
S_M=\int_\Omega d^4y\mathscr{L}_M\sqrt{-g},\ \
\mathscr{L}_M=\mathscr{L}_M(\psi, \mathring{\nabla}_a\psi, c.c., x^\alpha,
\mathring{\nabla}_ax^\alpha),
\end{equation}
where $\Omega$ is an arbitrary domain of the flat spacetime ${\cal M}_0$, $\{y^\mu\}$ is an arbitrary
coordinate system on $\Omega$, $g$ is the determinant of the Minkowski metric $g_{\mu\nu}$,
$\psi$ is the matter field, $\mathring{\nabla}_a$ is the metric-compatible and torsion-free derivative,
$a$ is an abstract index \cite{Wald,Liang}, which shows that the quantity is independent of coordinate
choice, and can be transformed into any tetrad or coordinate index by taking the corresponding component,
$c.c.$ denotes the complex conjugate, and $x^\alpha$ are inertial coordinates. The theory is assumed to
be Lorentz invariant, i.e., $S_M$ is invariant under the transformation:
\begin{equation}\label{LT}
\psi\rightarrow T(h)\psi,\ \ x^\alpha\rightarrow h^\alpha{}_\beta x^\beta,
\end{equation}
where $h=h^\alpha{}_\beta$ is an element of the Lorentz group $SO(1,3)$, and $T$ is the representation
of $SO(1,3)$ associated with the matter field $\psi$. Note that the Minkowski metric
\begin{equation}\label{g-mink}
g_{ab}=\eta_{\alpha\beta}(\mathring{\nabla}_ax^\alpha)(\mathring{\nabla}_bx^\beta),
\end{equation}
where $\eta_{\alpha\beta}={\rm diag}(-1,1,1,1)$. It is Lorentz invariant, so as $\sqrt{-g}$.
Also, the theory is supposed to be diffeomorphism invariant in the sense that $S_M$ is independent of
the choice of $\{y^\mu\}$ and invariant under the transformation:
\begin{equation}\label{diff}
\Omega\rightarrow\phi[\Omega],\ \ \psi\rightarrow \phi_*\psi,\ \ x^\alpha\rightarrow \phi_*x^\beta,
\end{equation}
where $\phi$ is a diffeomorphism, and $\phi_*$ denotes the pushforward by it: $(\phi_*\psi)(\phi x)=
\psi(x)$, $\forall x\in{\cal M}_0$. We give an example of such a theory: the Dirac Lagrangian
\begin{equation}\label{Dirac}
\mathscr{L}_M=-\frac12\mathbbm{i}(\overline{\psi}\gamma^a\mathring{\nabla}_a\psi-c.c.)
+\mathbbm{i}m\overline{\psi}\psi,
\end{equation}
where $\mathbbm{i}$ is the imaginary unit, $\gamma^a=\gamma^\alpha (\partial/\partial x^\alpha)^a$,
and $\gamma^\alpha$ are Dirac matrices.

The localization of the above theory is to replace $h$ in Eq. (\ref{LT}) by a function valued
at $SO(1,3)$. To achieve this, introduce a connection 1-form $\Gamma^\alpha{}_{\beta a}$ valued
at $so(1,3)$, i.e., subject to $\Gamma_{\alpha\beta a}=-\Gamma_{\beta\alpha a}$. Then modify
$\mathring{\nabla}_a\psi$ and $\mathring{\nabla}_ax^\alpha$ to be
\begin{equation}\label{Dpsi}
D_a\psi=\mathring{\nabla}_a\psi+T_\alpha{}^\beta\Gamma^\alpha{}_{\beta a}\psi,
\end{equation}
\begin{equation}\label{Dx}
D_ax^\alpha=\mathring{\nabla}_ax^\alpha+\Gamma^\alpha{}_{\beta a}x^\beta,
\end{equation}
where $T_\alpha{}^\beta$ are representations of the Lorentz generators. It can
be checked that $S_M$ is invariant under Eq. (\ref{LT}) together with the connection
transformation:
\begin{equation}\label{GammaT}
\Gamma^\alpha{}_{\beta a}\rightarrow
h^\alpha{}_\gamma\Gamma^\gamma{}_{\delta a}(h^{-1})^\delta{}_\beta
+h^\alpha{}_\gamma \mathring{\nabla}_a(h^{-1})^\gamma{}_\beta.
\end{equation}
Then we say that the theory is locally Lorentz invariant.
Moreover, $S_M$ is still independent of the choice
of $\{y^\mu\}$, and invariant under Eq. (\ref{diff}) together with $\Gamma^\alpha{}_{\beta a}
\rightarrow\phi_*\Gamma^\alpha{}_{\beta a}$. In this sense, we say that the theory is
still diffeomorphism invariant. Also note that the metric (\ref{g-mink}) is modified to be
\begin{equation}\label{g-curv}
g_{ab}=\eta_{\alpha\beta}(D_ax^\alpha)(D_bx^\beta),
\end{equation}
which is not necessarily flat. Accordingly, the $x^\alpha$ are no longer inertial coordinates.
They become the components of some vector field and may be called the local inertial coordinates
(LIC). The geometrical meaning of $\Gamma^\alpha{}_{\beta a}$ and $x^\alpha$ can be read off from
Eqs. (\ref{GammaT})--(\ref{g-curv}): $e^\alpha{}_a\equiv D_ax^\alpha$ is an orthonormal
co-tetrad field, and $\Gamma^\alpha{}_{\beta a}$ is just the spacetime connection which defines
a metric-compatible derivative $\nabla_a$ by $e^\alpha{}_b\nabla_ae_\beta{}^b=\Gamma^\alpha{}_{\beta a}$.

The last step of the construction of Lorentz gravity is to determine $\Gamma^\alpha{}_{\beta a}$
dynamically, i.e., to write down its action integral $S_G$, which may be defined as
\begin{equation}\label{SG}
S_G=\int_\Omega d^4y\mathscr{L}_G\sqrt{-g},\ \
\mathscr{L}_G=\mathscr{L}_G(x^\alpha, D_ax^\alpha, R^\alpha{}_{\beta ab}),
\end{equation}
where
\begin{equation}\label{R}
R^\alpha{}_{\beta ab}=d_a\Gamma^\alpha{}_{\beta b}+\Gamma^\alpha{}_{\gamma a}\wedge\Gamma^\gamma{}_{\beta b}
\end{equation}
is the curvature 2-form of $\Gamma^\alpha{}_{\beta a}$, and $d_a$ is the exterior derivative defined by,
e.g., $d_a\Gamma^\alpha{}_{\beta b}=2\mathring{\nabla}_{[a}\Gamma^\alpha{}_{|\beta| b]}=\mathring{\nabla}_a
\Gamma^\alpha{}_{\beta b}-\mathring{\nabla}_b\Gamma^\alpha{}_{\beta a}$.
The gravitational field equations are given by $V_\alpha{}^{\beta a}
\equiv\delta S/\delta\Gamma^\alpha{}_{\beta a}=0$ and $V_\alpha\equiv\delta S/\delta x^\alpha=0$, where
$S=S_M+S_G$. It follows from a direct computation that
\begin{equation}\label{AM}
V_\alpha{}^{\beta a}=\frac{\partial\mathscr{L}}{\partial D_a\psi}T_\alpha{}^\beta\psi
+c.c.+2D_b\frac{\partial\mathscr{L}}{\partial R^\alpha{}_{\beta ab}}
+\left(\frac{\partial\mathscr{L}}{\partial D_ax^{[\alpha}}+\mathscr{L}D^ax_{[\alpha}\right)\cdot x^{\beta]},
\end{equation}
\begin{equation}\label{Valpha}
V_\alpha=\frac{\partial\mathscr{L}}{\partial x^\alpha}-D_a\left(\frac{\partial\mathscr{L}}{\partial D_ax^\alpha}
+\mathscr{L}D^ax_\alpha\right),
\end{equation}
where $\mathscr{L}=\mathscr{L}_M+\mathscr{L}_G$, and $(D^ax_{[\alpha})\cdot x^{\beta]}=
(D^ax_{[\alpha})\cdot x_{\gamma]}\eta^{\gamma\beta}$.

\subsection{Noether's theorem in Lorentz gravity}
Now let us generalize Noether's theorem \cite{Noether} to Lorentz gravity, i.e., find out
the conservation laws corresponding to the Lorentz and diffeomorphism symmetries.
Summarizing the results in the last subsection, the action integral and Lagrangian function
of the coupling system of a matter field and a Lorentz gravitational field are as follows:
\begin{equation}
S=\int_\Omega d^4y\mathscr{L}\sqrt{-g},\ \
\mathscr{L}=\mathscr{L}(\psi, D_a\psi, c.c., x^\alpha, D_ax^\alpha, R^\alpha{}_{\beta ab}).
\end{equation}
The action $S$ is independent of $\{y^\mu\}$, and invariant under the transformation
\[
\Omega\rightarrow\phi[\Omega],\ \ \psi\rightarrow T(h)\phi_*\psi,
\ \ x^\alpha\rightarrow h^\alpha{}_\beta\phi_*x^\beta,
\]
\begin{equation}\label{SymT}
\Gamma^\alpha{}_{\beta a}\rightarrow
h^\alpha{}_\gamma(\phi_*\Gamma^\gamma{}_{\delta a})(h^{-1})^\delta{}_\beta
+h^\alpha{}_\gamma \mathring{\nabla}_a(h^{-1})^\gamma{}_\beta.
\end{equation}
To derive the conservation law, vary $\phi$ and $h$ to the one-parameter groups $\{\phi_t\}$
and $\{h_t\}$. Denote $(d/dt)|_{t=0}$ by $\delta$, then it follows from the chain rule that
\begin{eqnarray}\label{chain}
\delta\mathscr{L}=\frac{\partial\mathscr{L}}{\partial\psi}\delta\psi+
\frac{\partial\mathscr{L}}{\partial D_a\psi}\delta D_a\psi+c.c.+
\frac{\partial\mathscr{L}}{\partial x^\alpha}\delta x^\alpha\nonumber\\
+\frac{\partial\mathscr{L}}{\partial D_ax^\alpha}\delta D_ax^\alpha+
\frac{\partial\mathscr{L}}{\partial R^\alpha{}_{\beta ab}}\delta R^\alpha{}_{\beta ab}.
\end{eqnarray}
The variations $\delta\mathscr{L}$, $\delta\psi$, etc. can be expressed in terms of the generators
of $\{\phi_t\}$ and $\{h_t\}$, denoted by $v^a$ and $A^\alpha{}_\beta$.
The vector field $v^a$ at any point $p$ of the spacetime is equal to the tangent vector
of the curve $\phi_tp$, and the $so(1,3)$-valued function $A^\alpha{}_\beta=\delta h^\alpha{}_\beta$.
Note that the diffeomorphism in Eq. (\ref{SymT}) is gauge dependent, i.e., $\phi$ and $h$ do not commute.
It seems that a gauge-independent diffeomorphism is more natural to be a fundamental symmetry
transformation, which can be defined by Eq. (\ref{SymT}) with $A^\alpha{}_\beta=-\Gamma^\alpha{}_{\beta a}v^a$.
This transformation is interpreted as a translation in Poincar\'e gravity \cite{Hehl76}. As will
be shown later, this interpretation does not hold in the present framework.
Generally, set $A^\alpha{}_\beta=B^\alpha{}_\beta
-\Gamma^\alpha{}_{\beta a}v^a$, where $B^\alpha{}_\beta$ is an $so(1,3)$-valued function.
Then $B^\alpha{}_\beta$ stands for a Lorentz rotation, and $v^a$ a gauge-independent diffeomorphism.
Now it is ready to write down the variations $\delta\mathscr{L}$, $\delta\psi$, etc. in Eq. (\ref{chain}).
The result is: $\delta\mathscr{L}=-v^a\mathring{\nabla}_a\mathscr{L}$,
\[
\delta\psi=B^\alpha{}_\beta T_\alpha{}^\beta\psi-v^aD_a\psi,\ \
\delta x^\alpha=B^\alpha{}_\beta x^\beta-v^aD_ax^\alpha,
\]\[
\delta D_a\psi=B^\alpha{}_\beta T_\alpha{}^\beta D_a\psi-v^bD_bD_a\psi
-(D_b\psi)\mathring{\nabla}_av^b,
\]\[
\delta D_ax^\alpha=B^\alpha{}_\beta D_ax^\beta-v^bD_bD_ax^\alpha
-(D_bx^\alpha)\mathring{\nabla}_av^b,
\]
\begin{equation}\label{variation}
\delta R^\alpha{}_{\beta ab}=[B^\alpha{}_\gamma,R^\gamma{}_{\beta ab}]
-v^cD_cR^\alpha{}_{\beta ab}-R^\alpha{}_{\beta cb}\mathring{\nabla}_av^c
-R^\alpha{}_{\beta ac}\mathring{\nabla}_bv^c.
\end{equation}
Suppose that the matter field equation $\delta S/\delta\psi=0$ is satisfied, then substitution
of Eq. (\ref{variation}) into Eq. (\ref{chain}) leads to
\begin{eqnarray}\label{arb-v}
\mathring{\nabla}_b\mathscr{L}=
\left(D_a\frac{\partial\mathscr{L}}{\partial D_a\psi}\right)D_b\psi
+\frac{\partial\mathscr{L}}{\partial D_a\psi}D_bD_a\psi+c.c.
+\frac{\partial\mathscr{L}}{\partial x^\alpha}D_bx^\alpha\nonumber\\
+\frac{\partial\mathscr{L}}{\partial D_ax^\alpha}D_bD_ax^\alpha
+\frac{\partial\mathscr{L}}{\partial R^\alpha{}_{\beta ac}}D_bR^\alpha{}_{\beta ac},
\end{eqnarray}
\begin{equation}\label{arb-dv}
0=\frac{\partial\mathscr{L}}{\partial D_a\psi}D_b\psi+c.c.
+\frac{\partial\mathscr{L}}{\partial D_ax^\alpha}D_bx^\alpha
+2\frac{\partial\mathscr{L}}{\partial R^\alpha{}_{\beta ac}}R^\alpha{}_{\beta bc},
\end{equation}
\begin{eqnarray}\label{arb-B}
0=\left(D_a\frac{\partial\mathscr{L}}{\partial D_a\psi}\right)T_\alpha{}^\beta\psi
+\frac{\partial\mathscr{L}}{\partial D_a\psi}T_\alpha{}^\beta D_a\psi+c.c.
+\frac{\partial\mathscr{L}}{\partial x^{[\alpha}}x^{\beta]}\nonumber\\
+\frac{\partial\mathscr{L}}{\partial D_ax^{[\alpha}}D_ax^{\beta]}
+\frac{\partial\mathscr{L}}{\partial R^\alpha{}_{\gamma ab}}R^\beta{}_{\gamma ab}
-\frac{\partial\mathscr{L}}{\partial R^\gamma{}_{\beta ab}}R^\gamma{}_{\alpha ab},
\end{eqnarray}
where the arbitrariness of $v^a$, $\mathring{\nabla}_av^b$ and $B^\alpha{}_\beta$ at any given point is used.

The conservation law is just hidden in the identities (\ref{arb-v})--(\ref{arb-B}).
To see this, define the EM tensor $\Sigma_b{}^a=(\partial\mathscr{L}/\partial D_ax^\alpha)
D_bx^\alpha+\mathscr{L}\delta^a{}_b$, and the spin tensor $\tau_\alpha{}^{\beta a}=
(\partial\mathscr{L}/\partial D_a\psi)T_\alpha{}^\beta\psi+c.c.
+2D_b(\partial\mathscr{L}/\partial R^\alpha{}_{\beta ab})$. Then Eq. (\ref{arb-dv}) implies
that $\Sigma_b{}^a=-(\partial\mathscr{L}/\partial D_a\psi)D_b\psi+c.c.
-2(\partial\mathscr{L}/\partial R^\alpha{}_{\beta ac})R^\alpha{}_{\beta bc}+\mathscr{L}\delta^a{}_b$.
Also note that $[D_a,D_b]T=\sum_U\mathring{R}^c{}_{dab}T^d-\sum_LT_c\mathring{R}^c{}_{dab}
+\sum_U{\cal R}^i{}_{jab}T^j-\sum_LT_i{\cal R}^i{}_{jab}$, where $T$ is a tensor field valued
at some tensor space of $V_R$, $V_R$ is a representation space of $SO(1,3)$, $i,j$ are the
indices of $V_R$, $\sum_U$ denotes a sum for the upper indices of $T$, and $\sum_L$ for the
lower indices, $\mathring{R}^c{}_{dab}$ is the curvature tensor of $\mathring{\nabla}_a$,
${\cal R}^i{}_{jab}$ is the representation of $R^\alpha{}_{\beta ab}$, and the indices of $T$
are omitted except for those interacting with the curvature. It is also instructive to note
that $D_aT_\alpha{}^\beta\equiv\mathring{\nabla}_aT_\alpha{}^\beta+[\omega_a,T_\alpha{}
^\beta]=-2T_{[\alpha}{}^\gamma\Gamma^{\beta]}{}_{\gamma a}$,
$D_aD_bx^\alpha=K_b{}^\alpha{}_a$, and $D_{[c}R^\alpha{}_{|\beta|ab]}=0$,
where $\omega_a=\Gamma^\alpha{}_{\beta a}T_\alpha{}^\beta$,
$K^c{}_{ab}=(S^c{}_{ab}+S_{ab}{}^c+S_{ba}{}^c)/2$ is the contorsion tensor of $\nabla_a$,
and $S^\alpha{}_{ab}=d_ae^\alpha{}_b+\Gamma^\alpha{}_{\beta a}\wedge e^\beta{}_b$ is the torsion
2-form. With the help of the above definitions and formulas, Eqs. (\ref{arb-v}) and (\ref{arb-B})
can be reformulated into
\begin{equation}\label{EM}
\mathring{\nabla}_a\Sigma_b{}^a=-\Sigma_c{}^aK^c{}_{ab}-\tau_c{}^{da}R^c{}_{dab}
+\frac{\partial\mathscr{L}}{\partial x^\alpha}e^\alpha{}_b,
\end{equation}
\begin{equation}\label{spin}
D_a\tau_\alpha{}^{\beta a}=-\Sigma_{[\alpha}{}^{\beta]}
-\frac{\partial\mathscr{L}}{\partial x^{[\alpha}}x^{\beta]}.
\end{equation}
In the SR limit with $R^\alpha{}_{\beta ab}=0$ and so $S^\alpha{}_{ab}=0$, Eq.
(\ref{EM}) reduces to $\partial_a\Sigma_\alpha{}^a=\partial\mathscr{L}/\partial x^\alpha$,
which shows that the EM current is not conserved unless $\partial\mathscr{L}/\partial
x^\alpha=0$. Note that $\partial\mathscr{L}/\partial x^\alpha=0$ is not forced to hold by the
diffeomorphism invariance defined here. On the other hand, the AM current is
conserved. To prove this, first define the orbital AM current $\Sigma_\alpha{}^{\beta a}
=\Sigma_{[\alpha}{}^ax^{\beta]}$. Then from Eq. (\ref{AM}), $V_\alpha{}^{\beta a}
=\tau_\alpha{}^{\beta a}+\Sigma_\alpha{}^{\beta a}$ is just the total AM current.
Moreover, the combination of Eqs. (\ref{EM}) and (\ref{spin}) leads to $D_aV_\alpha{}^{\beta a}
=(\Sigma_c{}^aS^c{}_{ba}+\tau_c{}^{da}R^c{}_{dba})e^b{}_{[\alpha}x^{\beta]}$.
In the SR limit, this equation reduces to $\partial_aV_\alpha{}^{\beta a}=0$,
which is just the AM conservation law. Finally, according to Eqs. (\ref{Valpha}) and (\ref{EM}),
we have $V_\alpha=\partial\mathscr{L}/\partial x^\alpha-D_a\Sigma_\alpha{}^a
=(\Sigma_c{}^aS^c{}_{ab}+\tau_c{}^{da}R^c{}_{dab})e^b{}_\alpha$, and hence the AM conservation
law has the following elegant form:
\begin{equation}\label{AMcons}
D_aV_{\alpha\beta}{}^a=x_{[\alpha}V_{\beta]}.
\end{equation}
In the SR limit, $V_\alpha=0$ is automatically satisfied, and thus the $x^\alpha$ can be
viewed as an auxiliary field. In conclusion, the Lorentz and diffeomorphism symmetries
in Lorentz gravity only result in the AM conservation (\ref{AMcons}). This implies that
the diffeomorphism invariance defined here does not lead to the EM conservation,
and therefore it should not stand for the translational invariance.

\section{Poincar\'e gravity}
\subsection{Poincar\'e gravity from the gauge principle}
The Poincar\'e gravity is a gauge theory of the Poincar\'e group.
First consider a classical field theory with global Poincar\'e invariance and diffeomorphism
invariance. The action integral and Lagrangian function of this matter field are as follows:
\begin{equation}
S_M=\int_\Omega d^4y\mathscr{L}_M\sqrt{-g},\ \
\mathscr{L}_M=\mathscr{L}_M(\psi, \mathring{\nabla}_a\psi, c.c., \xi^A,
\mathring{\nabla}_a\xi^A),
\end{equation}
where $\xi^A=(\xi^\alpha,l)$, $\xi^\alpha$ are inertial coordinates of the flat spacetime
${\cal M}_0$, and $l$ is a constant with the dimension of length\footnote{In dS gravity,
the analog of $l$ is the radius of the internal dS space, which is linked to the cosmological
constant by $\Lambda=3/l^2$ \cite{Lu13}. But here $l$ is an arbitrary constant without physical
meaning.}, which may be seen as the 5th
coordinate of ${\cal M}_0$ as a plane embedded in the 5d flat space. The theory is assumed
to be Poincar\'e invariant, i.e., $S_M$ is invariant under the transformation:
\begin{equation}\label{PT}
\psi\rightarrow T(h)\psi,\ \ \xi^A\rightarrow H^A{}_B\,\xi^B,
\end{equation}
where $h=h^\alpha{}_\beta$ is an element of the Lorentz group $SO(1,3)$, $T$ is the
representation of $SO(1,3)$ associated with the matter field $\psi$, and $H^A{}_B$ is
a 5d representation of the Poincar\'e group, which satisfies $H^\alpha{}_\beta
=h^\alpha{}_\beta$, $H^4{}_\alpha=0$, and $H^4{}_4=1$, such that $H^A{}_B\,\xi^B=
(\xi'^\alpha,l)$ with $\xi'^\alpha=h^\alpha{}_\beta\xi^\beta+H^\alpha{}_4\cdot l$.
Note that the matter field is valued at a representation space of $SO(1,3)$, rather
than that of the full Poincar\'e group. We make this choice just for naturalness.
The reader who is interested at the latter kind of matter fields may refer to Ref.
\cite{Kazmierczak}. Notice that the Minkowski metric
\begin{equation}\label{g2}
g_{ab}=\eta_{AB}(\mathring{\nabla}_a\xi^A)(\mathring{\nabla}_b\xi^B)
\end{equation}
is Poincar\'e invariant, where $\eta_{AB}={\rm diag}(-1,1\cdots1)$. Also, the theory
is supposed to be diffeomorphism invariant in the sense that $S_M$ is independent of the
choice of $\{y^\mu\}$ and invariant under the transformation:
\begin{equation}\label{diff-P}
\Omega\rightarrow\phi[\Omega],\ \ \psi\rightarrow \phi_*\psi,\ \ \xi^A\rightarrow \phi_*\xi^A,
\end{equation}
where $\phi$ is a diffeomorphism. Recall that an example (\ref{Dirac}) is given for the
special theory of Lorentz gravity. As a matter of fact, the Dirac theory (\ref{Dirac}) is
also Poincar\'e invariant and diffeomorphism invariant in the above sense.

The localization of the above theory is to replace $H^A{}_B$ in Eq. (\ref{PT}) by a function valued
at the Poincar\'e group. To that end, introduce a connection 1-form $\Omega^A{}_{Ba}$ valued at the
Poincar\'e algebra, i.e., subject to the condition that $\Omega_{\alpha\beta a}=-\Omega_{\beta\alpha a}$
and $\Omega^4{}_{Ba}=0$. Then modify $\mathring{\nabla}_a\psi$ and $\mathring{\nabla}_a\xi^A$ to be
\begin{equation}
D_a\psi=\mathring{\nabla}_a\psi+T_\alpha{}^\beta\Omega^\alpha{}_{\beta a}\psi,
\end{equation}
\begin{equation}
D_a\xi^A=\mathring{\nabla}_a\xi^A+\Omega^A{}_{Ba}\xi^B,
\end{equation}
where $T_\alpha{}^\beta$ are representations of the Lorentz generators. It can
be checked that $S_M$ is invariant under Eq. (\ref{PT}) together with the connection
transformation:
\begin{equation}\label{Ommega-T}
\Omega^A{}_{Ba}\rightarrow H^A{}_C\,\Omega^C{}_{Da}(H^{-1})^D{}_B
+H^A{}_C\mathring{\nabla}_a(H^{-1})^C{}_B.
\end{equation}
Then we say that the theory is locally Poincar\'e invariant. Moreover, $S_M$ is still
independent of the choice of $\{y^\mu\}$, and invariant under Eq. (\ref{diff-P}) together
with $\Omega^A{}_{Ba}\rightarrow\phi_*\Omega^A{}_{Ba}$. In this sense we say that the
theory is still diffeomorphism invariant. Also note that the metric (\ref{g2}) is modified
to be \cite{Pilch,Lu13}
\begin{equation}\label{g2-curv}
g_{ab}=\eta_{AB}(D_a\xi^A)(D_b\xi^B),
\end{equation}
which is not necessarily flat. Accordingly, the $\xi^A$ are no longer inertial coordinates.
They become the components of some 5-vector field and may be called the (5d) LIC.
The geometrical meaning of $\Omega^A{}_{Ba}$ and $\xi^A$ can be read off from
Eqs. (\ref{Ommega-T})--(\ref{g2-curv}): $e^\alpha{}_a\equiv D_a\xi^\alpha$ is an orthonormal
co-tetrad field, and $\Omega^\alpha{}_{\beta a}$ is just the spacetime connection which defines
a metric-compatible derivative $\nabla_a$ by $e^\alpha{}_b\nabla_ae_\beta{}^b=\Omega^\alpha{}_{\beta a}$.
Notice that in the Lorentz gauge with $\xi^A=(0\cdots0,l)$, $D_a\xi^\alpha=\Omega^\alpha{}_{4a}\cdot l$
and so $\Omega^\alpha{}_{4a}=e^\alpha{}_a\cdot l^{-1}$.

The last step of the construction of Poincar\'e gravity is to determine $\Omega^A{}_{Ba}$
dynamically, i.e., to write down its action integral $S_G$, which may be defined as
\begin{equation}
S_G=\int_\Omega d^4y\mathscr{L}_G\sqrt{-g},\ \
\mathscr{L}_G=\mathscr{L}_G(\xi^A, D_a\xi^A, {\cal F}^A{}_{Bab}),
\end{equation}
where
\begin{equation}
{\cal F}^A{}_{Bab}=d_a\Omega^A{}_{Bb}+\Omega^A{}_{Ca}\wedge\Omega^C{}_{Bb}
\end{equation}
is the curvature 2-form of $\Omega^A{}_{Ba}$. It can be verified that ${\cal F}^\alpha{}_{\beta ab}
=R^\alpha{}_{\beta ab}$ is the Lorentz curvature (\ref{R}), and in the Lorentz gauge
${\cal F}^\alpha{}_{4ab}=S^\alpha{}_{ab}\cdot l^{-1}$ is the torsion 2-form. Finally,
the gravitational field equations are given by $V_A{}^{Ba}\equiv\delta S/\delta\Omega^A{}_{Ba}=0$
and $V_A\equiv\delta S/\delta\xi^A=0$, where $S=S_M+S_G$. On account of $\Omega^4{}_{Ba}=0$
and $\xi^4=l$, it follows that $V_4{}^{Ba}\equiv0$ and $V_4\equiv0$.
Moreover, it can be shown that
\begin{equation}\label{Valpha-a}
V_\alpha{}^{\beta a}=\frac{\partial\mathscr{L}}{\partial D_a\psi}T_\alpha{}^\beta\psi
+c.c.+2D_b\frac{\partial\mathscr{L}}{\partial{\cal F}^{[\alpha}{}_{\beta]ab}}
+\left(\frac{\partial\mathscr{L}}{\partial D_a\xi^{[\alpha}}+\mathscr{L}e_{[\alpha}{}^a
\right)\cdot\xi^{\beta]},
\end{equation}
\begin{equation}\label{V4-a}
V_\alpha{}^{4a}=2D_b\frac{\partial\mathscr{L}}{\partial{\cal F}^\alpha{}_{4ab}}
+\left(\frac{\partial\mathscr{L}}{\partial D_a\xi^\alpha}+\mathscr{L}e_\alpha{}^a\right)\cdot l,
\end{equation}
\begin{equation}\label{VA}
V_\alpha=\frac{\partial\mathscr{L}}{\partial\xi^\alpha}-D_a\left(\frac{\partial\mathscr{L}}{\partial D_a\xi^\alpha}
+\mathscr{L}e_\alpha{}^a\right),
\end{equation}
where $\mathscr{L}=\mathscr{L}_M+\mathscr{L}_G$, and $D_b(\partial\mathscr{L}/
\partial{\cal F}^{[\alpha}{}_{\beta]ab})=D_b(\partial\mathscr{L}/
\partial{\cal F}^{[\alpha}{}_{\gamma ab})\eta_{\delta]\gamma}\eta^{\delta\beta}$.
Note that although $\partial\mathscr{L}/\partial{\cal F}^\alpha{}_{\beta ab}$
is anti-symmetric about $\alpha$ and $\beta$, its Poincar\'e-covariant derivative
is not necessarily anti-symmetric.

\subsection{Noether's theorem in Poincar\'e gravity}
Now let us generalize Noether's theorem to Poincar\'e gravity, i.e., find out the
conservation laws corresponding to the Poincar\'e and diffeomorphism symmetries.
Summarizing the results in the last subsection, the action integral and Lagrangian
function of the coupling system of a matter field and a Poincar\'e gravitational
field are as follows:
\begin{equation}
S=\int_\Omega d^4y\mathscr{L}\sqrt{-g},\ \
\mathscr{L}=\mathscr{L}(\psi, D_a\psi, c.c., \xi^A, D_a\xi^A, {\cal F}^A{}_{Bab}).
\end{equation}
The action $S$ is independent of $\{y^\mu\}$, and invariant under the transformation
\[
\Omega\rightarrow\phi[\Omega],\ \ \psi\rightarrow T(h)\phi_*\psi,
\ \ \xi^A\rightarrow H^A{}_B\phi_*\xi^B,
\]
\begin{equation}\label{SymT-P}
\Omega^A{}_{Ba}\rightarrow
H^A{}_C(\phi_*\Omega^C{}_{Da})(H^{-1})^D{}_B
+H^A{}_C\mathring{\nabla}_a(H^{-1})^C{}_B.
\end{equation}
To derive the conservation law, vary $\phi$ and $H$ to the one-parameter groups $\{\phi_t\}$
and $\{H_t\}$. Then it follows from the chain rule that
\begin{eqnarray}\label{chain-P}
\delta\mathscr{L}=\frac{\partial\mathscr{L}}{\partial\psi}\delta\psi+
\frac{\partial\mathscr{L}}{\partial D_a\psi}\delta D_a\psi+c.c.+
\frac{\partial\mathscr{L}}{\partial \xi^A}\delta\xi^A\nonumber\\
+\frac{\partial\mathscr{L}}{\partial D_a\xi^A}\delta D_a\xi^A+
\frac{\partial\mathscr{L}}{\partial {\cal F}^A{}_{Bab}}\delta{\cal F}^A{}_{Bab}.
\end{eqnarray}
The variations $\delta\mathscr{L}$, $\delta\psi$, etc. can be expressed in terms of the generators
of $\{\phi_t\}$ and $\{H_t\}$, denoted by $v^a$ and $A^A{}_B$. As before, the diffeomorphism in Eq.
(\ref{SymT-P}) is gauge dependent, i.e., $\phi$ and $H$ do not commute. The gauge-independent
diffeomorphism can be defined by Eq. (\ref{SymT-P}) with $A^A{}_B=-\Omega^A{}_{Ba}v^a$.
Generally, set $A^A{}_B=B^A{}_B-\Omega^A{}_{Ba}v^a$, where $B^A{}_B$ is a function valued at
the Poincar\'e algebra. Then $B^A{}_B$ stands for a Poincar\'e transformation, and $v^a$ a gauge-independent
diffeomorphism. Now it is ready to write down the variations $\delta\mathscr{L}$, $\delta\psi$, etc.
in Eq. (\ref{chain-P}). The result is: $\delta\mathscr{L}=-v^a\mathring{\nabla}_a\mathscr{L}$,
\[
\delta\psi=B^\alpha{}_\beta T_\alpha{}^\beta\psi-v^aD_a\psi,\ \
\delta\xi^A=B^A{}_B\xi^B-v^aD_a\xi^A,
\]\[
\delta D_a\psi=B^\alpha{}_\beta T_\alpha{}^\beta D_a\psi-v^bD_bD_a\psi
-(D_b\psi)\mathring{\nabla}_av^b,
\]\[
\delta D_a\xi^A=B^A{}_BD_a\xi^B-v^bD_bD_a\xi^A
-(D_b\xi^A)\mathring{\nabla}_av^b,
\]
\begin{equation}\label{var-P}
\delta{\cal F}^A{}_{Bab}=[B^A{}_C,{\cal F}^C{}_{Bab}]
-v^cD_c{\cal F}^A{}_{Bab}-{\cal F}^A{}_{Bcb}\mathring{\nabla}_av^c
-{\cal F}^A{}_{Bac}\mathring{\nabla}_bv^c.
\end{equation}
Suppose that the matter field equation $\delta S/\delta\psi=0$ is satisfied, then substitution
of Eq. (\ref{var-P}) into Eq. (\ref{chain-P}) leads to
\begin{eqnarray}\label{arbP-v}
\mathring{\nabla}_b\mathscr{L}=
\left(D_a\frac{\partial\mathscr{L}}{\partial D_a\psi}\right)D_b\psi
+\frac{\partial\mathscr{L}}{\partial D_a\psi}D_bD_a\psi+c.c.
+\frac{\partial\mathscr{L}}{\partial \xi^A}D_b\xi^A\nonumber\\
+\frac{\partial\mathscr{L}}{\partial D_a\xi^A}D_bD_a\xi^A
+\frac{\partial\mathscr{L}}{\partial{\cal F}^A{}_{Bac}}D_b{\cal F}^A{}_{Bac},
\end{eqnarray}
\begin{equation}\label{arbP-dv}
0=\frac{\partial\mathscr{L}}{\partial D_a\psi}D_b\psi+c.c.
+\frac{\partial\mathscr{L}}{\partial D_a\xi^A}D_b\xi^A
+2\frac{\partial\mathscr{L}}{\partial{\cal F}^A{}_{Bac}}{\cal F}^A{}_{B bc},
\end{equation}
\[
0=\left(D_a\frac{\partial\mathscr{L}}{\partial D_a\psi}\right)T_\alpha{}^\beta\psi
+\frac{\partial\mathscr{L}}{\partial D_a\psi}T_\alpha{}^\beta D_a\psi+c.c.
+\frac{\partial\mathscr{L}}{\partial\xi^{[\alpha}}\xi^{\beta]}
\]
\begin{equation}\label{arbP-B}
+\frac{\partial\mathscr{L}}{\partial D_a\xi^{[\alpha}}D_a\xi^{\beta]}
+\frac{\partial\mathscr{L}}{\partial{\cal F}^\alpha{}_{\gamma ab}}{\cal F}^\beta{}_{\gamma ab}
-\frac{\partial\mathscr{L}}{\partial{\cal F}^\gamma{}_{\beta ab}}{\cal F}^\gamma{}_{\alpha ab}
+\frac{\partial\mathscr{L}}{\partial{\cal F}^{[\alpha}{}_{4ab}}{\cal F}^{\beta]}{}_{4ab},
\end{equation}
\begin{equation}\label{arb-B4}
0=\frac{\partial\mathscr{L}}{\partial\xi^\alpha}\cdot l-
\frac{\partial\mathscr{L}}{\partial{\cal F}^\beta{}_{4ab}}{\cal F}^\beta{}_{\alpha ab},
\end{equation}
where the arbitrariness of $v^a$, $\mathring{\nabla}_av^b$, $B^\alpha{}_\beta$ and
$B^\alpha{}_4$ at any given point is used.

The conservation law is hidden in the identities (\ref{arbP-v})--(\ref{arb-B4}).
To see this, define the orbital EM tensor $\Sigma_b{}^a=(\partial\mathscr{L}
/\partial D_a\xi^A)D_b\xi^A+\mathscr{L}\delta^a{}_b$, and the Poincar\'e spin current
$\tau_A{}^{Ba}$: $\tau_\alpha{}^{\beta a}=(\partial\mathscr{L}/\partial D_a\psi)
T_\alpha{}^\beta\psi+c.c.+2D_b(\partial\mathscr{L}/\partial{\cal F}^{[\alpha}{}_{\beta]ab})$,
$\tau_\alpha{}^{4a}=2D_b(\partial\mathscr{L}/\partial{\cal F}^\alpha{}_{4ab})$, $\tau_4{}^{Aa}=0$.
Then Eq. (\ref{arbP-dv}) implies that $\Sigma_b{}^a=-(\partial\mathscr{L}/\partial D_a\psi)D_b\psi+c.c.
-2(\partial\mathscr{L}/\partial{\cal F}^A{}_{Bac}){\cal F}^A{}_{Bbc}+\mathscr{L}\delta^a{}_b$.
Also note that $[D_a,D_b]T=\sum_U\mathring{R}^c{}_{dab}T^d-\sum_LT_c\mathring{R}^c{}_{dab}
+\sum_U{\cal F}^i{}_{jab}T^j-\sum_LT_i{\cal F}^i{}_{jab}$, where $T$ is a tensor field valued
at some tensor space of $V_R$, $V_R$ is a representation space of the Poincar\'e group, $i,j$
are the indices of $V_R$, and ${\cal F}^i{}_{jab}$ is the representation of ${\cal F}^A{}_{Bab}$.
It is also instructive to note that $D_aT_\alpha{}^\beta=-2T_{[\alpha}{}^\gamma\Omega^{\beta]}{}_{\gamma a}$, $D_aD_b\xi^A=(K_b{}^\alpha{}_a,0)$, and $D_{[c}{\cal F}^A{}_{|B|ab]}=0$. With the help of the above definitions
and formulas, Eqs. (\ref{arbP-v}), (\ref{arbP-B}) and (\ref{arb-B4}) can be reformulated into
\begin{equation}\label{dsigma}
\mathring{\nabla}_a\Sigma_b{}^a=-\Sigma_c{}^aK^c{}_{ab}-\tau_A{}^{Ba}{\cal F}^A{}_{Bab}
+\frac{\partial\mathscr{L}}{\partial\xi^A}D_b\xi^A,
\end{equation}
\begin{equation}\label{dtau-alpha}
D_a\tau_{[\alpha}{}^{\beta]a}=-\Sigma_{[\alpha}{}^{\beta]}
-\frac{\partial\mathscr{L}}{\partial\xi^{[\alpha}}\xi^{\beta]},
\end{equation}
\begin{equation}\label{dtau4}
D_a\tau_\alpha{}^{4a}=-\frac{\partial\mathscr{L}}{\partial\xi^\alpha}\cdot l.
\end{equation}
Furthermore, define the 5d orbital AM current $\Sigma_A{}^{Ba}$: $\Sigma_\alpha{}^{\beta a}
=\Sigma_{[\alpha}{}^a\xi^{\beta]}$, $\Sigma_\alpha{}^{4a}=\Sigma_\alpha{}^a\cdot l$, and
$\Sigma_4{}^{Aa}=0$. Then from Eqs. (\ref{Valpha-a})--(\ref{V4-a}), $V_A{}^{Ba}=\tau_A{}^{Ba}
+\Sigma_A{}^{Ba}$ is just the total AM current. The combination of Eqs. (\ref{dsigma}) and
(\ref{dtau-alpha}) leads to $D_aV_{[\alpha}{}^{\beta]a}=(\Sigma_c{}^aS^c{}_{ba}+\tau_A{}^{Ba}
{\cal F}^A{}_{Bba})e^b{}_{[\alpha}\xi^{\beta]}$.
In the SR limit with ${\cal F}^A{}_{Bab}=0$, this equation reduces to $\partial_aV_\alpha{}
^{\beta a}=0$, which is just the AM conservation. Moreover, the combination of Eqs.
(\ref{dsigma}) and (\ref{dtau4}) yields $D_aV_\alpha{}^{4a}=(\Sigma_c{}^aS^c{}_{\alpha a}+
\tau_A{}^{Ba}{\cal F}^A{}_{B\alpha a})\cdot l$. In the SR limit, this equation becomes
$\partial_a\Sigma_\alpha{}^a=0$, which is just the EM conservation. Finally, according
to Eqs. (\ref{VA}) and (\ref{dsigma}), we have $V_\alpha=\partial\mathscr{L}/\partial\xi^
\alpha-D_a\Sigma_\alpha{}^a=\Sigma_c{}^aS^c{}_{a\alpha}+\tau_A{}^{Ba}{\cal F}^A{}_{Ba\alpha}
=V_A{}^{Ba}{\cal F}^A{}_{Ba\alpha}$, and so
the AM and EM conservation have the following elegant form:
\begin{equation}\label{AMC}
D_aV_{[\alpha}{}^{\beta]a}=-V_{[\alpha}\cdot\xi^{\beta]},
\end{equation}
\begin{equation}\label{EMC}
D_aV_\alpha{}^{4a}=-V_\alpha\cdot l.
\end{equation}
In the SR limit, $V_\alpha=0$ is automatically satisfied. In the Poincar\'e gravity, $V_\alpha=0$
as long as $V_A{}^{Ba}=0$. As a result, the $\xi^A$ can be viewed as an auxiliary field in both cases.
Generally, two gauge invariants can be defined from $V_A{}^{Ba}$ and $V_A$: the EM tensor $V_b{}^a=
V_\alpha{}^{4a}D_b\xi^\alpha\cdot l^{-1}$, and $V_a=V_\alpha D_a\xi^\alpha$. Note that $V_A{}^{Ba}$
is a vector field valued at the dual space of the Poincar\'e algebra, and hence $V_\alpha{}^{\beta a}$
is not Poincar\'e covariant. But in the Lorentz gauge,
$V_\alpha{}^{\beta a}=\tau_\alpha{}^{\beta a}$ is Lorentz covariant, and so a Lorentz-invariant spin
tensor can be defined: $\tau_c{}^{da}=\tau_\alpha{}^{\beta a}e^\alpha{}_ce_\beta{}^d$. Then the
conservation laws (\ref{AMC})--(\ref{EMC}) reduce to $\overline{D}_a\tau_{\alpha\beta}{}^a=-V_{[\alpha
\beta]}$ and $\mathring{\nabla}_aV_b{}^a=-V_c{}^aK^c{}_{ab}-\tau_c{}^{da}R^c{}_{dab}$, where $\overline
{D}_a$ is the Lorentz-covariant derivative, and $V_{\alpha\beta}=V_{ba}e_\alpha{}^be_\beta{}^a$.
These two equations are the AM and EM conservation in the ordinary form \cite{Kibble,Hehl76,Hehl95}.
In conclusion, Eqs. (\ref{AMC})--(\ref{EMC}) generalize the ordinary conservation laws from the
Lorentz gauge to the general gauge.

\section{Remarks}
It can be concluded from our analysis that the diffeomorphism invariance defined in the
LIC formalism is different from the translational invariance. In the LIC formalism, the
translational invariance is treated as an internal symmetry. This makes the gravitational theory
look more like a matter gauge theory. Also, the LIC formalism presents a unified framework for
the Poincar\'e and dS gravity, such that both the Poincar\'e and dS invariance are independent
of the diffeomorphism invariance.

Besides from the standard tetrad formalism and the LIC formalism presented here, there also exist
the nonlinear realization \cite{Tseytlin,Tresguerres}/Cartan geometry \cite{Wise} formalisms for
the Poincar\'e gravity. In these formalisms, the dynamical variables are the nonlinear connection/Cartan
connection, which project the linear connection into the Lorentz gauges. These formalisms may be useful
for the situation with symmetry
breaking. But in this paper, the Poincar\'e gravity is discussed in the Lagrangian level, in which
there exists no symmetry breaking. Then the LIC formalism, with the dynamical variable being the linear
connection, becomes a better choice.

Finally, I would like to give a remark on the choice of the
gravitational gauge group. If the AM and EM conservation are required, the gauge group
should be the dS/AdS/Poincar\'e group. Comparing the dS/AdS gravity and Poincar\'e
gravity, I find that the former is more elegant because the AM and EM currents can be
unified there. Actually, the AM and EM conservation (\ref{AMC})--(\ref{EMC}) can be
combined into a neat form as $D_aV_{AB}{}^a=\xi_{[A}V_{B]}$ in the dS/AdS gravity
\cite{Lu16}.

\section*{Acknowledgments}
The project is funded by the China Postdoctoral Science Foundation
under Grant No. 2015M572393, and the Fundamental Research Funds for the Central
Universities (Grant No. 161gpy49) at Sun Yat-sen University.

\end{document}